\documentclass{emulateapj}
\usepackage{amssymb}
\usepackage{apjfonts}
\usepackage{epsfig}
\usepackage{natbib}
\bibpunct{(}{)}{;}{a}{}{,}

\newcommand{\msun}{\mbox{$M_{\odot}$}}

\def\deg      {{\ifmmode^\circ\else$^\circ$\fi}} %%% Overwrites TeX \deg

\def\lsim{\mathrel{\rlap{\lower4pt\hbox{\hskip1pt$\sim$}}\raise1pt\hbox{$<$}}}
\def\gsim{\mathrel{\rlap{\lower4pt\hbox{\hskip1pt$\sim$}}\raise1pt\hbox{$>$}}}   
\slugcomment{}

\shorttitle{Star formation and dust obscuration at z$\approx$2}
\shortauthors{Pannella et al.}

\begin{document}


\title{Star formation and dust obscuration at $z\approx2$:
galaxies at the dawn of downsizing\altaffilmark{*}}
%\title{Star formation and dust obscuration at z$\approx$2: galaxies at the dawn of downsizing}


\author{M.~Pannella\altaffilmark{1},~C.~L.~Carilli\altaffilmark{1},~E.~Daddi\altaffilmark{2},~H.~J.~Mc~Cracken\altaffilmark{3},~F.~N.~Owen\altaffilmark{1},~A.~Renzini\altaffilmark{4},~V.~Strazzullo\altaffilmark{1},\\~F.~Civano\altaffilmark{5},~A.~M.~Koekemoer\altaffilmark{6},~E.~Schinnerer\altaffilmark{7},~N.~Scoville\altaffilmark{8},~V.~Smol\v{c}i\'{c}\altaffilmark{8},~Y.~Taniguchi\altaffilmark{9},~H.~Aussel\altaffilmark{2},\\~J.~P.~Kneib\altaffilmark{10},~O.~Ilbert\altaffilmark{11,12},~Y.~Mellier\altaffilmark{3},~M.~Salvato\altaffilmark{8},~D.~Thompson\altaffilmark{13},~C.~J.~Willott\altaffilmark{14}.}

\altaffiltext{*}{Based on observations collected, within the COSMOS Legacy Survey, at the HST, Chandra, XMM, Keck, NRAO-VLA, Subaru, KPNO, CTIO, CFHT and ESO observatories. The National Radio Astronomy Observatory is a
facility of the National Science Foundation operated under cooperative
agreement by Associated Universities, Inc.}
\altaffiltext{1}{National Radio Astronomy Observatory, P.O. Box 0, Socorro, NM 87801-0387;~mpannell@nrao.edu~.}
\altaffiltext{2}{CEA, Laboratoire AIM - CNRS - Universit\'e Paris Diderot,  
Irfu/SAp, Orme des Merisiers, F-91191 Gif-sur-Yvette, France.}
\altaffiltext{3}{Institut d'Astrophysique de Paris, 98 bis Boulevard Arago, 75014 Paris, France.}
\altaffiltext{4}{INAF - Osservatorio Astronomico di Padova, Vicolo dell'Osservatorio 5, I-35122 Padova, Italy.}
\altaffiltext{5}{Harvard Smithsonian Center for Astrophysics
60 Garden Street, MS 67
Cambridge, MA 02138.}
\altaffiltext{6}{Space Telescope Science Institute 3700 San Martin Drive, Baltimore MD 21218.}
\altaffiltext{7}{Max Planck Institut f¨ur Astronomie, K¨onigstuhl 17, Heidelberg, D-69117, Germany.}
\altaffiltext{8}{California Institute of Technology, MS 105-24, Pasadena, CA 91125.}
\altaffiltext{9}{Graduate School of Science and Engineering, Ehime University, Bunkyo-cho, Matsuyama 790-8577, Japan.}
\altaffiltext{10}{Laboratoire d'Astrophysique de Marseille, Technopôle de Marseille-Etoile 38, rue Frédéric Joliot-Curie 13388 Marseille cedex 13 FRANCE}
\altaffiltext{11}{Institute for Astronomy, 2680 Woodlawn Dr.,  
University of Hawaii, Honolulu, Hawaii, 96822}
\altaffiltext{12}{Laboratoire d'Astrophysique de Marseille, BP 8,  
Traverse du Siphon, 13376 Marseille Cedex 12, France}
\altaffiltext{13}{Large Binocular Telescope Observatory U.of.A, 933 N. Cherry   Ave.Tucson, AZ}
\altaffiltext{14}{Herzberg Institute of Astrophysics, National Research Council, 5071 West Saanich Rd, Victoria, BC V9E 2E7, Canada}


\begin{abstract}
We present first results of a study aimed to constrain the star
formation rate and dust content of galaxies at z$\approx$2.  We use a
sample of BzK-selected star-forming galaxies, drawn from the COSMOS survey, to perform a
stacking analysis of their 1.4 GHz radio continuum as a function of
different stellar population properties, after removing
AGN contaminants from the sample. Dust unbiased star formation rates 
are derived from radio fluxes assuming the local radio-IR
correlation. The main results of this work are: i) specific star formation rates are constant 
over about 1 dex in stellar mass and up to the highest stellar mass probed; ii) the dust attenuation is a strong function of galaxy stellar mass with more massive galaxies being more obscured
than lower mass objects; iii) a single value of the UV extinction applied to all galaxies would lead to grossly
underestimate the SFR in massive galaxies; iv)~correcting the observed UV luminosities for dust attenuation based  on the Calzetti recipe provide results in very good agreement with the radio derived ones; v) the mean specific star formation rate of our sample steadily decreases by a factor of $\sim 4$ with decreasing redshift from $z=2.3$ to 1.4 and a factor of $\sim 40$ down the local Universe. 
 
These empirical SFRs would cause galaxies to dramatically overgrow in mass if maintained all the way to low redshifts, we suggest that this does not happen because star formation is progressively quenched, likely starting from the most massive galaxies. 

\end{abstract}





\section{Introduction}

How and when galaxies build up their stellar mass is still a major
question in observational cosmology. While a general consensus has
been reached in the last years on the evolution of the galaxy stellar
mass function~\citep[e.g.][]{dickinson2003,drory2004,bundy2005,P06,font06,marchesini2008},
the redshift evolution of the star formation rate as a function of
stellar mass SFR$(M,z)$ still remains unclear. Several studies, mainly
based on UV derived SFR, have found
the persistence of the, locally well established, anticorrelation between 
specific star formation rate and stellar mass up to high redshift~\citep[e.g.][]{juneau2005,feulner05,bauer2005,erb2006,noeske07,zheng2007,cowie2008,damen2009,davies2009}. This anticorrelation is often regarded in the literature as a manifestation of the
"downsizing" scerario~\citep{cowie1996}, whereby
more massive galaxies form at higher redshift. Part of this effect is
certainly real, as for $z<2$ the most massive galaxies tend to be
passively evolving ellipticals with no or little ongoing star
formation. However, when referring to actively star forming galaxies alone whether
this anticorrelation exists or not depends on the way SFRs are estimated. As
already warned by Cowie et al. (1996), one important and poorly known
ingredient in deriving SFRs from
rest-frame UV fluxes is the amount of dust attenuation suffered
by the UV light in the inter-stellar medium. Lacking spectral information for large samples, the 
dust attenuation factor is the result of a multi-parameter, and highly degenerate, fitting to the multiwavelength photometry available 
or, when the photometric coverage is not sufficient, a median factor is applied to the whole galaxy sample.  

An independent estimate of the star formation rate in a galaxy, not biased by the galaxy's
dust content, is provided by its radio continuum emission. This is due to processes, the free-free emission from HII regions and the synchrotron radiation from relativistic electrons, dominated by young massive stars. By mean of the well established (but not as well understood) radio-FIR
correlation~\citep[e.g.][]{condon1992,kennicutt1998,yun2001} it is possible to estimate the total star formation rate in a galaxy from its radio luminosity. Thanks to their arcsecond resolution and relatively wide field of
view, radio interferometric observations offer several advantages over
present-day FIR facilities which are limited by their $\sim10''$
resolution and narrow field of view.  For this very reason radio
continuum observations turn out to be an excellent tool for tracing the
dust-unobscured star formation in the high redshift Universe.

However, radio emission is not only produced by star formation but
also by AGN, and therefore a major challenge in deriving dust-unbiased
SFRs from radio fluxes is to remove the AGN contamination~\citep[e.g.][]{smolcic2008}.  
At $z>1$, even in the deepest present-day surveys, radio detections are likely to include
a substantial population of AGNs, although extreme
ULIRG/SMG--like starbursting galaxies do exist. Therefore the best way to
explore, with existing radio facilities, the dust-unbiased SFRs of
normal galaxy populations is to use a stacking analysis of the radio
data, which allows the investigation of large galaxy samples drawn from
optical-NIR surveys that are individually undetected in the radio. 
This technique has been already used in a number of
radio studies \citep[e.g.][]{daddi071, white07, carilli2007, carilli08,dunne2008, garn2008}.

In this context the Cosmic Evolution Survey (COSMOS, \citealt{sco2007}), with its state-of-the-art multiwavelength coverage all the
way from X-rays to radio of a 2$\sq\degr$ field provides an ideal
opportunity to build large high redshift galaxy samples with well
characterized spectral properties. We take advantage of the COSMOS database
to select a large sample of $1<z<3$ star-forming galaxies, and derive dust-unbiased SFRs from stacking the 1.4 GHz radio data.

Throughout this paper we use AB magnitudes and adopt a $\Lambda$
cosmology with \mbox{$\Omega_M=0.3$}, \mbox{$\Omega_\Lambda=0.7$} and
\mbox{$H_0=70 \; \mathrm{km} \, \mathrm{s}^{-1} \, \mathrm{Mpc}^{-1}$}.

\section{RADIO, OPTICAL, NEAR INFRARED AND X-RAY DATA} 
\label{sec_data}

Here we use the VLA medium-deep 1.4GHz imaging covering the
whole COSMOS field with a fairly uniform rms
($\approx$10$\mu$Jy) and an angular resolution of 1.5$"$ (see \citealt{schinnerer2007}). 

Deep SUBARU $B,z$ imaging (\citealt{capak2007}) and CFHT $K_{\rm s}$-band
data (McCracken et al. 2009, submitted) were used to select a
dust-unbiased sample of about 34,000 star-forming BzK galaxies (sBzK,
see left panel of Figure~1) with $K_{\rm s}<23$. We refer to McCracken et al for a detailed description 
of the $K$-selected BzK sample.  Following
~\cite{daddi2004} galaxy stellar masses were estimated assuming a \cite{salpeter1955} initial mass function from 0.1 to 100$M_\odot$. A photometric redshift was assigned to more
than 80\% of the sBzK sample, by cross-correlating with the COSMOS
photometric redshift catalog by \cite{ilbert2009}. The median
photo-$z$ is $\sim 1.7$, with less than 2\% of the sample at redshift
lower than 1 or higher than 3 (see right panel of Figure~1), confirming the
effectiveness of the BzK selection technique.

\begin{figure}
  \includegraphics[height=.24\textwidth]{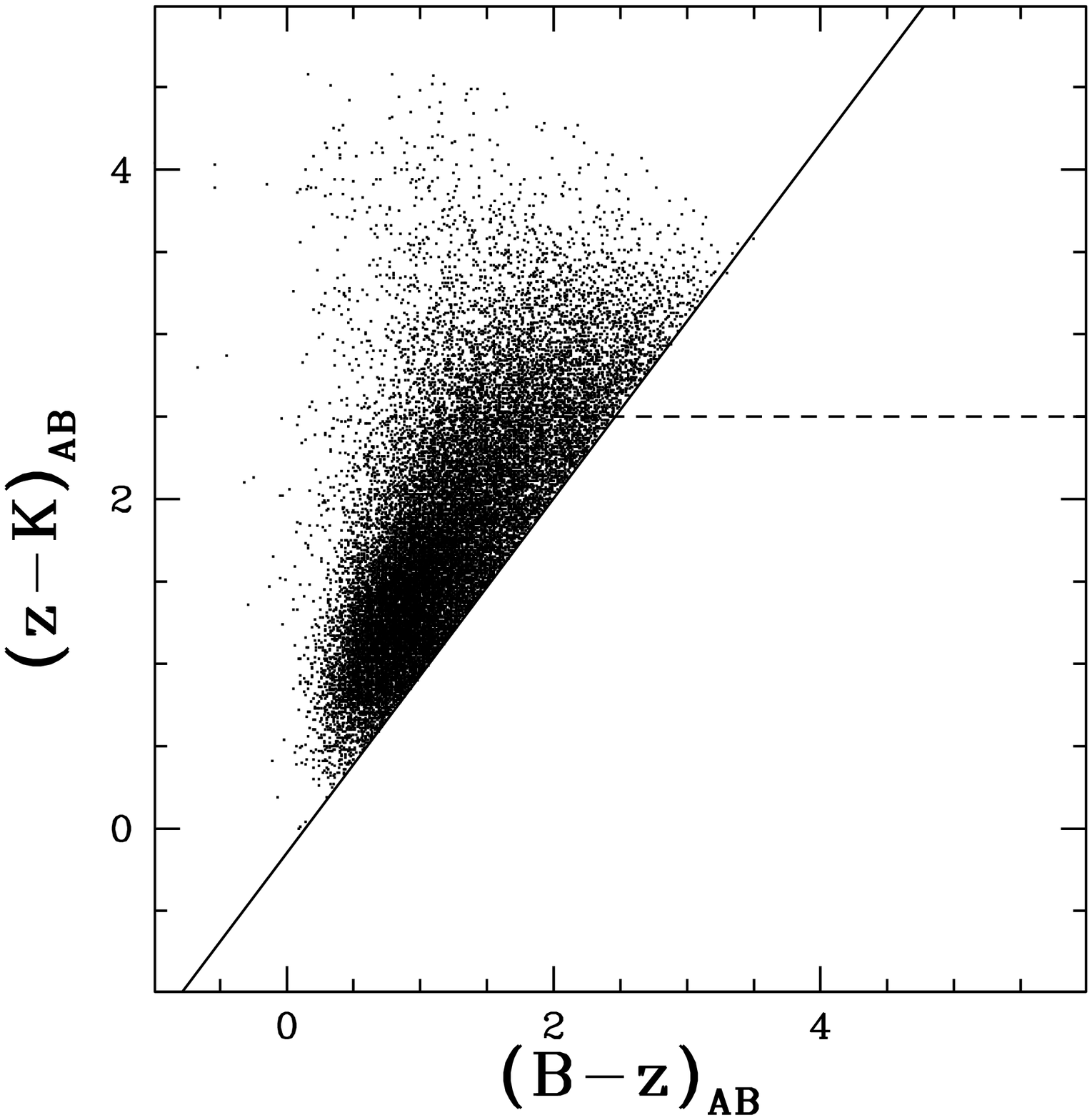}%
  \includegraphics[height=.24\textwidth]{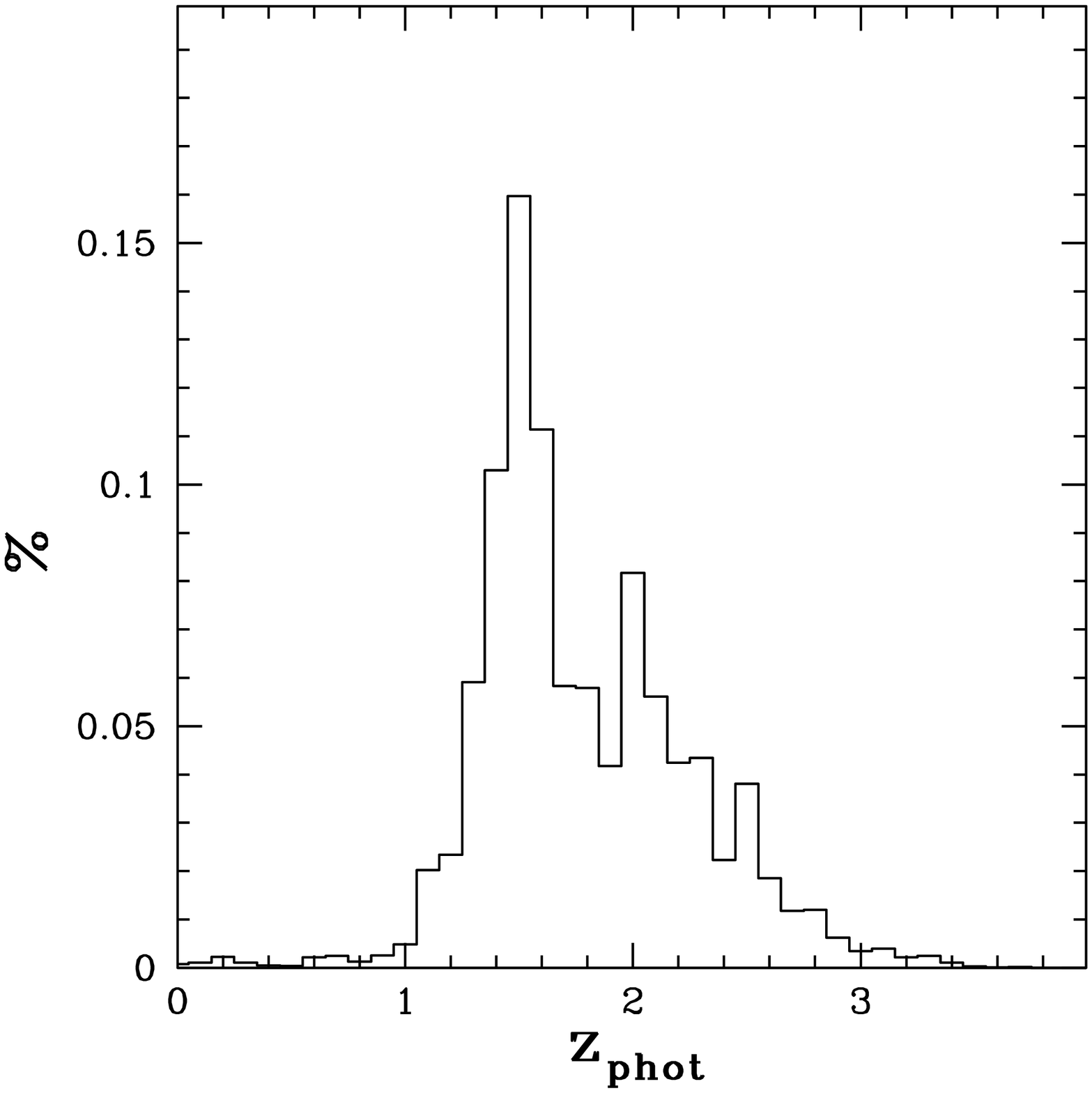}
  \caption{{\bf Left:} The selection diagram for sBzK star forming
  galaxies at z$\approx$2. {\bf Right:} Photometric redshift
  distribution of the COSMOS sBzK sample. The bulk of the sample
  spans the redshift range [1.3-2.5].}
\end{figure}

A small fraction ($\approx$2\%) of the sBzK sample has a 1.4GHz
counterpart. The minimum flux density of the radio counterparts, corresponding to a 3$\sigma$ detection, is about 30$\mu$Jy which, at a median redshift of 1.7, corresponds to a
radio luminosity of about 5$\times$10$^{23}$W/Hz at 1.4GHz.
  
In the local Universe it is usually assumed, based on radio luminosity function studies~\citep[e.g.][]{sadler2002,condon2002}, that 1.4GHz radio luminosities greater than $\approx$2$\times$10$^{23}$~W/Hz are mostly produced by AGNs, while below this luminosity star formation has a dominant role in producing the observed radio emission, maybe still in concurrence with a low-luminosity AGN. Even though recent studies (e.g. Smol{\v c}i{\'c} et al. 2009ab; Strazzullo et al., in preparation) are suggesting that such a characteristic luminosity was brighter at higher redshift, the sBzK radio detections are mostly {\it extreme} objects: AGN dominated galaxies or SMG--like starbursts.

In order to
study star formation and dust content for the {\it normal} galaxy
population, mostly undetected in radio, we removed from the sample all
the objects with a radio counterparts. In doing so we are likely
removing, along with AGN dominated sources, also the tail of extreme
star forming objects. Nonetheless we prefer this conservative approach
in order to derive more robust conclusions. 

\begin{figure}[htbp!]
  \includegraphics[height=.16\textwidth, bb= 37 329 577 509,clip]{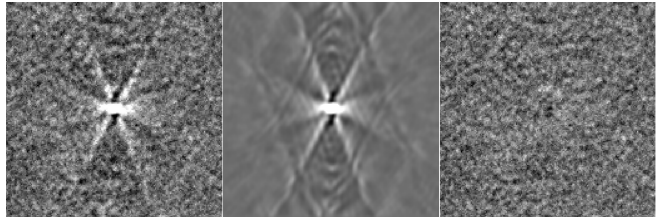}
  \caption{{\bf Left:}~Median stacking result of all the 34000 sBzK galaxies.~{\bf Middle:}~Best fit dirty beam
convolved Gaussian to the stacked data. The total flux recovered is
8.8$\pm$0.1$\mu$Jy. ~{\bf Right:}~Residual image. \label{fig:stack}}
\end{figure}

In a further attempt to
remove AGN contributed radio emission, we restrict our analysis to the
inner central 0.9 square degree of the COSMOS field, which is covered
by deep Chandra observations~\citep{elvis2009}. The depth of the Chandra survey reaches
$1.9\times10^{-16}$ erg cm$^2$ s$^{-1}$, in the soft band (0.5-2keV), which
at the median redshift of our sample allows an important census of the
AGN luminosity function. We cross-correlated the sBzK sample with the catalog of Chandra counterparts~(Civano et al. 2009, submitted), removing all
matched sources (575) from the final catalog.

\begin{figure*}
\begin{center}
  \includegraphics[height=.265\textwidth, bb = 30 160 565 370, clip]{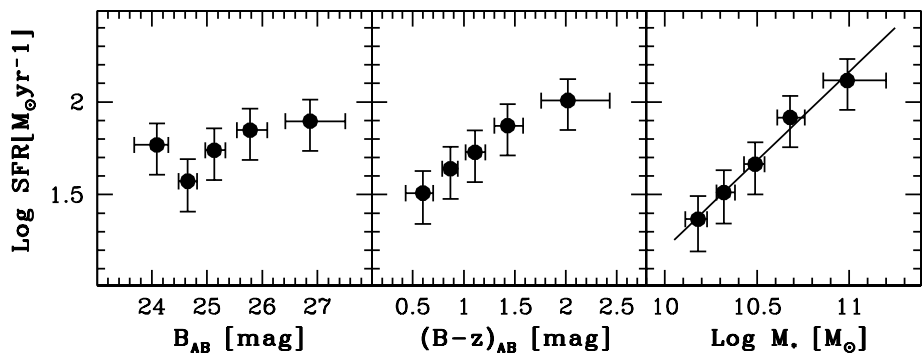}
  \caption{Total radio-derived SFR versus B band~(left), B-z
  color~(middle), and stellar mass~(right). The solid line is the best fit line: Log(SFR)$\propto$0.95~LogM$_*$.\label{fig:totalfluxes}}
\end{center}
\end{figure*}

Excluding individually-detected X-ray sources may not completely
eliminate AGN sources from the sample. Indeed, sBzK galaxies with a
mid-IR excess ($\sim 20-30\%$ of the total) are likely to contain
heavily obscured (Compton thick) AGNs, as indicated by their X-ray
stacking (Daddi et al. 2007b). However, mid-IR excess and non-excess
(normal) galaxies exhibit quite similar 1.4 GHz radio properties
(Daddi et al. 2007a), suggesting that such Compton-thick AGNs do not
contribute substantially to the radio flux at 1.4 GHz. Moreover, our
median stacking technique automatically reduces the impact of a
minority of galaxies with radio emission in excess from what is
expected from star formation, if they exist.

We ended up with a reduced sample of 11798 objects, over an area of $\approx$ 0.9 square degrees, with a photometric redshift in the range $1< z < 3$ and having removed both the radio 
(248) and the Chandra (575) detections. In the following we will present results based on this sub-sample but even analyzing the full sBzK sample our conclusions would remain substantially the same.

\section{Data analysis and Radio-Derived SFR}
\label{mmll}

For each of these sBzK sources, we produced a cutout in the radio
mosaic of 173$\times$173 pixel$^2$ (60.5 $\times$60.5
arcsec$^2$). These cutouts were then stacked to create median
images. Median stacking is more robust than mean against the tails of
the distribution, while the rms still goes down by $\approx\sqrt N$.
Stacking images allows an easy way to treat the
Bandwidth Smearing (BWS) effect. This is a well known instrumental
effect in aperture synthesis astronomy~\citep[see][]{thompson1999}, consisting of a
radial stretching of the sources in the image due to the finite width
of the frequency response of the receiver. The BWS does not affect the
total flux of the source though. Total fluxes are retrieved by fitting a dirty beam convolved with a Gaussian
function to the stacked data. The Galfit code~\citep{peng2002} was
used for this purpose, but very similar results were obtained using
the AIPS/CLEAN algorithm. As an example, in
Figure~\ref{fig:stack} we show the stacked data, model and residual
image for the original whole (34,000 galaxies) sBzK sample.
Given the large number of objects in our sample, we were able to stack
the radio continuum in bins of different galaxy properties, such as
magnitude, color, and mass. Measured radio fluxes were converted to
star formation rates using the median redshift of each stacked sample (1.7 for the
whole population), a synchrotron emission spectral index of --0.8, and
the conversion factor between radio luminosity and SFR from
\cite{yun2001}, {\it i.e.}
\begin{equation}
\rm SFR = 5.9 \pm 1.8 \times 10^{-22}~{\sl L}_{\rm 1.4GHz} \; (M_\odot /yr),
\end{equation}
\noindent where $L_{\rm 1.4\rm GHz}$ is in W Hz$^{-1}$. Errors on SFRs are the squared sum of the uncertainties coming from 
the off-source rms in the stacked images, the fitting to recover total fluxes, 
and the uncertainty in equation~(1).

In Figure~\ref{fig:totalfluxes} we show our results for the
radio stacking of the AGN--cleaned sBzK sample as a function of: {\it
i)} the observed $B$-band magnitude, which is related to the restframe
dust uncorrected UV luminosity; {\it ii)} the $(B-z)$ color, which for
galaxies at $z\sim 2$ is a proxy for the UV slope of the spectral
energy distribution and hence it relates to dust extinction; and {\it iii)}
the galaxy stellar mass. We conclude that: 1) overall, the emerging UV
light is poorly correlated with the ongoing SFR, and --somewhat
counter intuitively-- the highest SFRs are found among the UV-faintest
galaxies; 2) this happens because galaxies with higher SFRs are more
extinguished in the UV; and 3) the SFR increases with stellar mass almost
linearly, as the  slope of the Log(SFR)-Log $M_*$ relation  is
0.95$\pm$0.07, in agreement  (within the small errors) with  the relation
found by \citet{daddi071}.

\subsection{The specific star formation rate}
\label{ssfr}

\begin{figure*}
\begin{center}
  \includegraphics[height=.35\textwidth, bb = 25 150 530 450, clip]{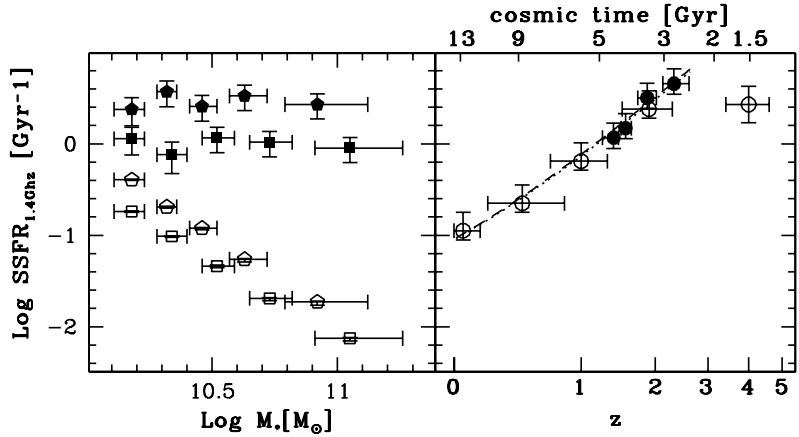}
 \caption{{\bf Left:} Radio derived SSFR (solid symbols) at z$\approx$ 1.6~(squares)~and $\approx$ 2.1~(pentagons)~are compared to
the uncorrected UV derived SSFR (empty symbols) as a function of Log
M$_*$. {\bf Right:} Radio derived SSFRs from this work~(solid dots), for star forming galaxies with M$_*\sim$~3$\times$10$^{10}$M$_{\odot}$, as a function of redshift at
z~$\approx$ 1.4, 1.6, 1.9, 2.3. Literature data are plotted as empty circles. The dotted curve shows the SSFR as a function of redshift described by equation~(2). 
 \label{fig:ssfr}}
\end{center}
\end{figure*}

In Figure~\ref{fig:ssfr} (left panel) we present radio derived specific
star formation rates (SSFR=SFR/$M_*$) for the reduced sBzK sample, divided in two redshift bins centered at z$\approx$1.6~(solid squares) and 2.1~(solid pentagons). From the observed $B$-band magnitudes we also derive UV$_{1500}$ luminosities, uncorrected for dust attenuation, then estimating an uncorrected UV--derived SSFRs, which are also plotted
in the same Figure with empty symbols. Some striking features are worth noting in the
plot: {\it i)} the UV-derived SSFR drops dramatically with increasing
mass whereas dust free SSFRs show no such effect, the SSFR being
constant over almost one dex in mass (see also \cite{dunne2008}); {\it
ii)} correcting the UV light with a single value of extinction
$A_{1500}$ at all masses (an approximation often adopted in the
literature, see e.g. ~\citealt{gabasch2004, juneau2005,bauer2005}) would result in an artificial decreasing SSFR with increasing mass; and {\it iii)} the mean dust attenuation is a function of the galaxy stellar mass, with more massive galaxies being more dust-extinguished.

By taking advantage of the available photometric redshifts, we can
split our sBzK sample in four redshift bins centered at
z$\approx$[1.4, 1.6, 1.9, 2.3] and look for the redshift evolution of
the SSFR, which is almost independent of stellar mass.  On the right
panel of Figure~4 we show how SSFRs are steadily increasing with
redshift, by a factor $\sim 4$ in the  explored redshift range. 

We also overplot three lower redshifts realizations~\citep{brinchmann04,noeske07,elbaz07} and the $z\sim2$ estimate by Daddi et al. (2007a), by computing the SSFR predicted from these studies for star forming galaxies with M$_*\sim$3$\times$10$^{10}$M$_{\odot}$, and show how the SSFRs have decreased by a factor 40, for this mass galaxies, from z$\approx$2.3 all the way down to the local 
Universe.

\subsection{The dust attenuation at 1500~\AA}
\label{dust}
By forcing the dust-corrected UV-SFRs to agree with the radio-SFRs, as
both a function of galaxy stellar mass and $(B-z)$ color, we obtain how the UV light 
attenuation A$_{1500}$ at $z\sim2$ relates to these quantities. The result is shown in the inserts of Figure 5. 
\cite{meurer1999} found a similar relation for a sample of
local starburst galaxies. Our relation naturally extends their results
to higher redshifts, and also nicely shows that the sBzK selection is much less
biased against highly obscured objects than UV-selected samples.
The latter ones are indeed limited to moderate extinctions, such as  
$A_{1500}<$3.6 mag \citep{meurer1999}. 

In the explored redshift interval the dust attenuation, stellar
mass and SFR are all tightly correlated with each other. The left panel of Figure 5 shows that the dust extinction $A_{1500}$ tightly correlates
with galaxy mass. Therefore, assuming a constant value for
$A_{1500}$ (independent of galaxy mass) introduces a systematic bias and the resulting
SSFR($M_*)$ relation decreases with increasing stellar mass.

We emphasize the excellent agreement of the 
dust-attenuation correction here derived using the radio data with that derived from
the UV continuum slope: the dotted line in the right panel of Figure 5 shows the relation between attenuation and $(B-z)$ color predicted by the Calzetti et al. (1994) law, as calibrated in Daddi et al. (2004).

\begin{figure*}
\begin{center}
  \includegraphics[height=.32\textwidth, bb = 30 160 570 460, clip]{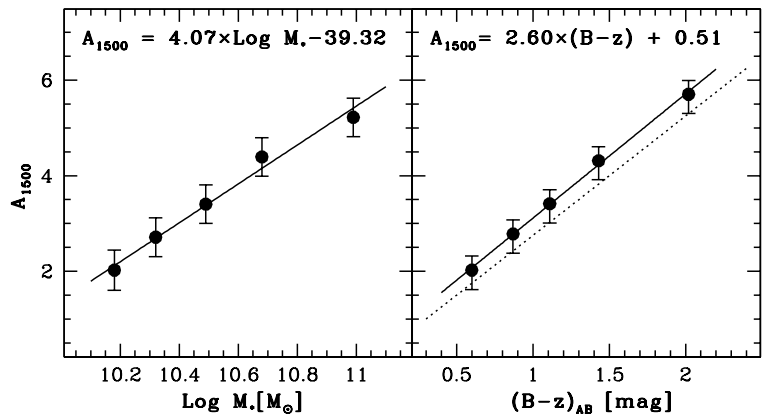}
  \caption{{\bf Left:}~UV light attenuation (A$_{1500}$ = 2.5$\times$Log(SFR$_{1.4GHz}$/SFR$_{1500}$) as a function of galaxy
  stellar mass.~{\bf Right:}~UV light attenuation as a function of B-z
  color~(UV slope). The dotted line shows the attenuation law derived
  in~\cite{daddi2004} as described in the text.
  \label{fig:AUV}}
\end{center}
\end{figure*}

\subsection{The Mass growth of galaxies}
\label{mass}

The present results confirm that, within the explored mass range, the
SSFR of $z\sim 2$ star-forming galaxies is almost independent of
stellar mass (Daddi et al. 2007a; Dunne et al. 2009).  A tight
correlation between SFR and stellar mass was also found at $z\sim 1$
(Elbaz et al. 2007), $z = 0.2-0.7$ (Noeske et al. 2007) and $z\sim 0$
(Brinchmann et al. (2004). 
Other studies extensively discussed by Dunne et al. (2009) find instead a
SSFR that declines appreciably with increasing stellar mass (see also Cowie \& Barger 2008). 
In this respect, we concur
with the arguments put forward by Dunne et al., that appear to be
strengthened by our findings.

 The SSFR secular decline and the mentioned results can be represented roughly by the relation:
\begin{equation}
\overline{\rm SFR}\simeq 270\, (M_*/10^{11}M_\odot)\, (t/3.4\times  10^9{\rm yr})^{-2.5} \;(M_\odot /yr),
\label{equa}
\end{equation}
where $t$ is the cosmic time.

We stress here that the exponent of $M_*$ in equation (2) may not be
strictly 1, and may depend on redshift (cf. Dunne et al. 2009), hence
this relation is best valid for $M_*\sim 3\times 10^{10}\msun$ and
$z<2.4$ ($T>2.7$ Gyr) for which if was derived. Still it represents a fair approximation for the star
forming galaxy population .~In Figure~4 we show that such a relation does not hold for the $z\sim 4$ galaxies belonging to the
~\citet{daddi2009} sample, this suggests that it has an important flattening above a certain redshift, qualitatively resembling in its behaviour the redshift evolution of the cosmic star formation history. 

Integrating equation (2) from z=2 to z=0, {\it i.e.} assuming that individual galaxies continued making stars and growing in mass all the way to low redshift, they would increase in mass by a factor $\sim$250, a clear overgrow even neglecting the contribution of merging events.

On the other hand, star-forming galaxies {\it do} form stars
at these high rates, so either equation~(2) is grossly erroneous, or at some point
it ceases to  apply to individual galaxies. 
%Given the nice agreement between UV-corrected and radio-derived SFRs, we
%tend to favor the second option. 
Indeed, between $z\sim 2.4$ and $z\sim
0$ a major transformation takes place in the population of galaxies.
While at $z\sim 2.4$ only a small fraction of the stellar
mass is in passively evolving galaxies (elliptical and bulges), this
fraction grows up to $\sim 60\%$ by $z=0$ ~\citep{baldry2004}. 
Therefore, we argue that equation~(2) does indeed apply to star-forming galaxies all the way to $z=0$, but star
formation turns off in a growing fraction of galaxies, which
progressively turn into passive ellipticals and bulges. Mapping
quantitatively this transformation  goes beyond the scope of the
present Letter.
%, and is one of the major goals of the ongoing COSMOS project.

\section{Conclusions}
\label{conc}
We have presented first results of a study aimed to investigate the
dust-unbiased star formation properties of high redshift galaxies, by
focusing on their stacked radio properties. We use a sample of
sBzK galaxies, drawn from the COSMOS survey, with a median
redshift of 1.7, rejecting known AGN identified in both deep  X-ray Chandra
data and the 1.4 GHz radio imaging.

%The highest SFRs occur in galaxies with the
%aintest UV luminosities, as galaxies with higher SFRs are more UV
%extinguished. We also find that dust attenuation is a function of galaxy
%stellar mass, with more massive galaxies being more heavily extinguished.
%As the observed UV luminosity is poorly correlated with the ongoing star
%formation, 

We demonstrate that a universal dust-attenuation correction cannot be applied to our sample. For instance, the generic factor of 5 often used to correct the UV light of Lyman-break galaxies (LBG) is applicable in our sample only for objects
with $M_*\sim 3\times 10^{10}\; M_\odot$ --which incidentally is very
close to the median stellar mass of LBG galaxies~\citep{shapley2001}--
but would grossly underestimate the correction for more massive
galaxies.  

We extend the results of Daddi et al.~(2007a) that UV light, appropriately corrected, is a reliably tracer of SFR at $z\sim2$.

We find that the SFR of star-forming galaxies increases almost linearly with stellar mass at all explored redshifts. 

It appears that we are witnessing an evolution era when almost all star forming galaxies had the same evolutionary timescales and a nearly exponential growth, independent of mass. This is consistent with Dunne et al.~(2009) results. They argue that the descrepancy found with literature studies might be due to selection biases present in UV and optically selected studies.  While we agree with their statement, we point out that an underestimate of the dust attenuation correction could also explain such discrepancy.  

We also find that the mass--independent SSFRs decrease by a factor 4 in
the redshift range from $z=2.3$ to 1.4, a trend that continues all the way to the local Universe. Individual galaxies would enormously overgrow in mass if these empirical SFRs were maintained down to low redshifts. 
We suggest that this does not happen because many galaxies turn passive, and
do so in a {\it downsized} fashion, because massive galaxies are first to
reach unsustainable SFR levels. Thus, in massive starforming galaxies at $z\sim 2$ downsizing has not started yet, but it will soon: we are just at the {\it dawn of downsizing}. 

Constraining the nature of the physical processes by which SSFR are kept
approximately constant in star forming galaxies of wildly different mass,
and the mechanisms that contribute to discontinuing the star formation 
activity in massive high redshift galaxies, are both substantial challenges
for theoretical models to reproduce and for observers to investigate 
in full detail. New ideas on gas accretion modes~\citep{dekel2009} and recent 
observations of widespread large molecular gas reservoirs~\citep{daddi08, tacconi2008} in distant 
massive galaxies will likely provide crucial paths to understand these issues.

\acknowledgments We thank the anonymous referee for constructive comments which improved the presentation of our results. MP, VS and CLC acknowledge partial support from the Max-Planck~Forschungspreise~2005. 
ED and HJMcC acknowledge support from the French grants ANR-07-BLAN-0228-03
and ANR-08-JCJC-0008. AR acknowlesges support from the ASI grant COFIS. 
This work is based in part on data products  produced at TERAPIX.
The HST COSMOS Treasury program was supported through NASA grant
HST-GO-09822. We gratefully acknowledge the contributions of the
entire COSMOS collaboration.

\end{document}